\documentclass{article}
\usepackage{spconf,amsmath,graphicx}
\usepackage{lipsum}
\usepackage{multicol}
\usepackage{booktabs}       
\usepackage{longtable}
\newcommand*{\thead}[1]{%
\multicolumn{1}{c}{\bfseries\begin{tabular}{@{}c@{}}#1\end{tabular}}}

\title{Multi-modal Medical Image Fusion For Non-Small Cell Lung Cancer Classification}
\name{Salma Hassan, Hamad Al Hammadi, Ibrahim Mohammed, Muhammad Haris Khan}
\address{Mohamed Bin Zayed University of Artificial Intelligence, Abu Dhabi, UAE}
\begin{document}
\maketitle

\begin{abstract}
The early detection and nuanced subtype classification of non-small cell lung cancer (NSCLC), a predominant cause of cancer mortality worldwide, is a critical and complex issue. In this paper, we introduce an innovative integration of multi-modal data, synthesizing fused medical imaging (CT and PET scans) with clinical health records and genomic data. This unique fusion methodology leverages advanced machine learning models, notably MedClip and BEiT, for sophisticated image feature extraction, setting a new standard in computational oncology. Our research surpasses existing 
approaches, as evidenced by a substantial enhancement in NSCLC detection and classification precision. The results showcase notable improvements across key performance metrics, including accuracy, precision, recall, and F1-score. Specifically, our leading multi-modal classifier model records an impressive accuracy of 94.04\%. We believe that our approach has the potential to transform NSCLC diagnostics, facilitating earlier detection and more effective treatment planning and, ultimately, leading to superior patient outcomes in lung cancer care.
\end{abstract}
\section{Introduction}
\label{sec:intro}

Lung cancer remains a significant health challenge, ranking as the second most prevalent cancer and the foremost cause of cancer-related mortality in both men and women \cite{WHO}. Characteristically, lung cancer symptoms often remain undetected until the disease has progressed to an advanced stage, complicating treatment efforts. In this context, image screening emerges as a critical tool, particularly for asymptomatic individuals. Studies underscore the efficacy of imaging techniques such as CT and MRI in early lung cancer detection among high-risk groups, including smokers and individuals with genetic predispositions, significantly enhancing survival prospects \cite{dodia2022recent,naseer2023lung}. However, existing methods for NSCLC classification, such as those utilizing standalone CT or PET imaging \cite{chaunzwa2021deep}, often face limitations in terms of sensitivity and specificity, and they may not fully capture the complexity of tumor heterogeneity. Additionally, existing approaches may lack the integration of comprehensive clinical and genetic data, which are crucial for a more precise diagnosis. These limitations underscore the need for more advanced diagnostic methodologies that can accurately pinpoint and characterize cancerous tissues, providing a strong motivation for our innovative multi-modal image fusion approach.

This paper aims to refine lung cancer diagnosis, particularly for non-small cell lung cancer (NSCLC), by integrating multi-modal data beyond traditional imaging scans. Our proposed method fuses CT and PET scans to leverage the strengths of both modalities. CT scans provide detailed anatomical information, while PET scans offer insights into metabolic activity, often indicative of tumor presence. By combining these scans, our fusion method creates a more comprehensive image that captures both structural and functional aspects of the lung. Such fusion techniques have shown promise in improving diagnostic accuracy in other organ scans and in combining CT with MRI \cite{dodia2022recent}.

Our approach addresses the limitations of existing diagnostic techniques that rely solely on single-modality imaging. CT or PET scans alone may not sufficiently differentiate between malignant and benign lesions, leading to diagnostic inaccuracies. Additionally, the standalone use of these modalities may only partially capture the tumor's complexity and heterogeneity. Therefore, our method begins with the denoising of scans using a deep CNN auto-encoder for more precise image reconstruction. This step is followed by the fusion of CT and PET scans through wavelet decomposition and accurate image registration to create a fused image that provides a richer, dual perspective of both functional and anatomical information. Specifically, fusing PET and CT images facilitates the precise localization of abnormal metabolic activities within the lung's structural framework \cite{amini2021multi}. This precise localization is crucial for identifying and characterizing cancerous tissues accurately. To complement this, we undertake rigorous pre-processing of both tabular and genetic data, encompassing missing value estimation, encoding, class balancing, scaling, and feature selection. The image processing phase includes optimal contrast enhancement and normalization.

By integrating multi-modal imaging with advanced machine learning techniques, we aim to enhance the overall diagnostic performance, leading to earlier and more accurate detection of NSCLC, ultimately improving patient outcomes. Leveraging advanced models like MedClip and BEiT, we enhance feature extraction and diagnostic accuracy. Additionally, incorporating clinical health records and genetic data allows for a more personalized diagnosis and treatment plan. This integration not only improves the sensitivity and specificity of NSCLC detection but also paves the way for earlier diagnosis and better patient outcomes.

This paper makes several key contributions to the field of oncological diagnostics, particularly in lung cancer:
\begin{itemize}
\item \textbf{Innovative Multi-Modal Data Fusion:} Introduces an advanced multi-modal data fusion approach that blends CT and PET imaging with clinical and genetic data. This methodology, relatively unexplored in lung cancer diagnostics, offers a more comprehensive diagnostic view, potentially leading to earlier and more accurate detection of NSCLC.
\item \textbf{Novel Application of Deep Denoising CNN Auto-Encoders:} Presents a novel application of deep CNN auto-encoders for the denoising of medical images, setting a new precedent for image clarity and diagnostic precision.
\item \textbf{Effective Integration of Diverse Data Types:} Demonstrates the effectiveness of integrating multiple data types through the use of diverse and sophisticated analytical models, significantly enhancing diagnostic accuracy.
\end{itemize}

Ultimately, the paper's core goal is to substantially improve the precision of NSCLC diagnosis by merging cutting-edge medical imaging with exhaustive patient data. By harnessing the complementary strengths of CT and PET scans, we aspire to generate a comprehensive, unified image that delineates lung anomalies more effectively. The fusion of this imaging data with clinical and genetic information forms the cornerstone of our NSCLC classification strategy, aiming to transform and refine diagnostic procedures in oncology.

\section{Related Work}
\label{sec:format}
Recent advancements in AI for diagnosing and treating NSCLC have significantly transformed patient outcomes in oncology \cite{alduais2023non}. Recent advancements in AI for diagnosing and treating NSCLC have significantly transformed patient outcomes in oncology \cite{alduais2023non}. The application of AI in NSCLC diagnosis encompasses several key research domains: medical imaging analysis, survival prediction, recurrence analytics, and multi-modal data integration. These fields have been instrumental in advancing the diagnosis and treatment of NSCLC. 

In the realm of medical imaging analysis, significant strides have been made in tumor detection and classification using AI algorithms applied to CT scans, as evidenced by the works of Feng and Khoirunnisa \cite{feng2022deep,khoirunnisa2023implementation}. These advancements have facilitated earlier and more precise diagnoses, consequently improving patient outcomes. Additionally, AI analysis of patient data has enhanced survival and recurrence prediction, offering valuable insights into patient prognosis \cite{zhang2015relationship}.

Recent research has also explored novel AI architectures for improving detection accuracy. Studies have investigated the potential of fully automated pipelines and advanced model architectures, such as CRNN, ViTs, AlexNet, and 3D reconstruction techniques, in enhancing diagnostic accuracy \cite{dodia2022recent, naseer2023lung}. However, the highest accuracy achieved to date in these studies was 84.1\% using the VGG16 model on medical imaging data alone \cite{han2021histologic}.

The integration of multi-modal data represents a burgeoning area of research. This includes both image multi-modality, combining MRI and CT scans, and data multi-modality, integrating tabular and imaging data \cite{feng2022deep,khoirunnisa2023implementation}. Studies by Feng have underscored the potential of combining imaging modalities like CT and MRI scans with patient genomics data to enhance diagnostic accuracy. The fusion of diverse data types, such as imaging modalities with patient genomic data, has shown promise in improving diagnostic accuracy. Our paper builds upon these foundational studies, aiming to pioneer the combination of multi-modal fused imaging with clinical and genomic data for NSCLC subtype classification. No work combines multi-modal fused imaging with clinical data and genomics sequences for the classification of both NSCLC subtypes, which is the scope and aim of this paper. 

Our paper contributes to this growing body of knowledge by not only applying these established techniques but also innovating in the way we integrate and process multi-modal data. The fusion of CT and PET scans, as proposed in our methodology, builds upon the foundational work of \cite{paudyal2023artificial}, who illustrated the benefits of CT and MRI image fusion in enhancing the clarity and informational value of medical scans for oncological application. Furthermore, our use of advanced models like MedClip and BEiT for image feature extraction extends the work of \cite{wang2022medclip}, who highlighted the strengths of these models in handling high-resolution medical images.

In summary, this paper stands at the intersection of several key research areas within the field of medical imaging and cancer diagnosis. By synthesizing these diverse methodologies and building upon them, our work aims to contribute a novel approach to the early detection and classification of NSCLC, addressing some of the limitations identified in previous studies.

\section{Proposed Methodology}
\label{sec:pagestyle}
Our methodology introduces several novel contributions to the field of NSCLC diagnosis through multi-modal data integration and advanced deep learning techniques. These contributions can be categorized into four key areas: comprehensive pre-processing, advanced image fusion,  innovative model architectures, and holistic data integration.

\begin{figure*}[ht]
  \includegraphics[width=\textwidth]{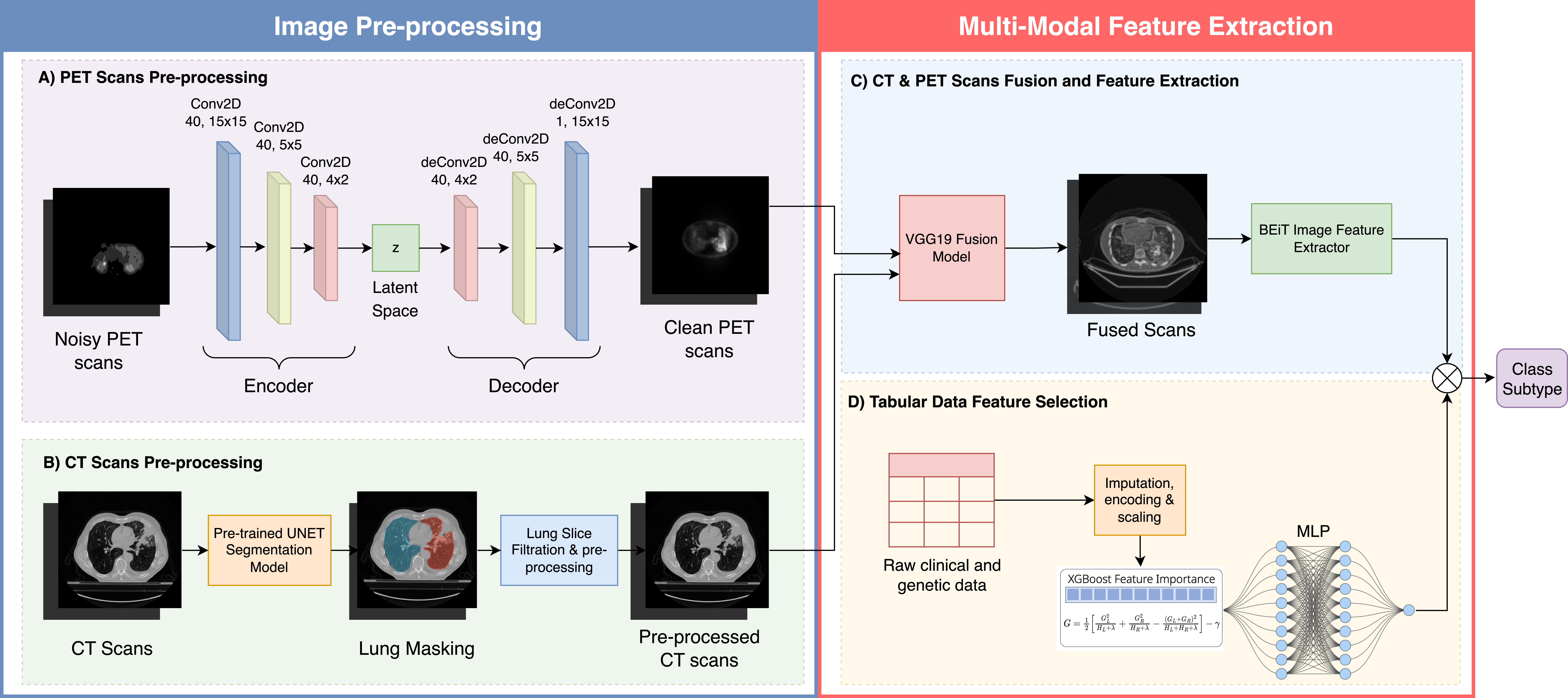}
  \caption{Multi-modal BEiT model showing model architecture that starts with pre-processing the CT and PET scan and then fusing them into a single scan that is then fused with the other data modalities, such as clinical and genetic, before applying feature selection and passing it to the BEiT model for classification.}
  \label{beit}
\end{figure*}

\subsection{Comprehensive Pre-processing}
Our approach to data pre-processing is a significant contribution with practical benefits. We address four key data modalities: clinical tabular data, genetic data, CT scans, and PET scans. For tabular and genetic data, we implement a series of rigorous pre-processing steps, including missing value imputation, categorical encoding, class balancing using SMOTE, and standardization. Feature importance is evaluated using XGBoost, ensuring that only the most relevant features are used in subsequent analyses. For imaging data, we develop a deep CNN auto-encoder trained on a clean PET scan dataset to de-noise PET scans, effectively reducing noise and enhancing image quality. CT scans undergo normalization and contrast enhancement. We also employ 3D Slicer software for precise image registration of CT and PET scans, ensuring accurate anatomical and functional alignment. These pre-processing steps are crucial for maximizing the performance of our fusion and classification models, leading to more accurate diagnoses and treatment plans in healthcare analytics.

\subsection{Advanced Image Fusion}
\begin{figure}[!ht]
    \centering
    \includegraphics[width=0.5\textwidth]{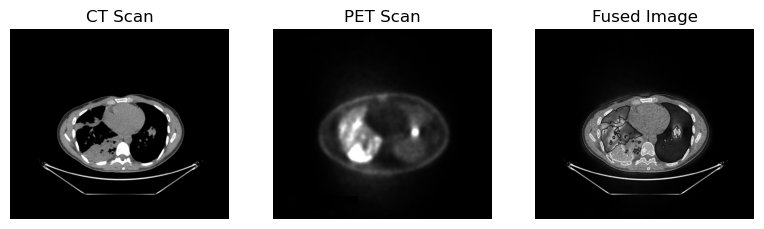}
    \caption{Example of CT, PET images, and resulting fused image}
    \label{fig:scans}
\end{figure}

One of the primary contributions of our methodology is the development of an advanced image fusion technique that combines the strengths of CT and PET scans. After all the prepossessing of the scans was done, the new clean input was fed into our VGG19 fusion model. The fusion algorithm decomposes the CT and PET scans into four coefficients, each using the discrete wavelet transform. The coefficients are the coefficient LL1 and three detail coefficients: LH1(horizontal), LV1(vertical), LD1(diagonal). After the fusion of the four pairs, inverse wavelet transform was applied to the four bands to obtain the fused image, results shown in Figure \ref{fig:scans}. The CT scan offers clear structural details of the lung tissues and surrounding areas, but it lacks metabolic information. The PET scan reveals areas of high metabolic activity indicative of tumor presence, but it lacks precise anatomical context. By merging these two modalities, the fused image not only overcomes the limitations of each modality but also provides a comprehensive view that is invaluable in medical imaging, highlighting both the detailed structures of the lung tissues and the areas of high metabolic activity. 

\subsection{Baseline Models}
The next stage involved implementing different baseline models to evaluate the effectiveness of the multi-modal approach. SVM and logistic regression were chosen for their robustness and simplicity in handling tabular and genetic data, providing reliable benchmarks for structured data analysis. For image classification, we selected 2D CNN, 3D CNN, VGG16, ResNet18, Inception, and Xception due to their proven efficacy in medical imaging tasks and their diverse architectures. These models represent a wide range of complexity and feature extraction capabilities. Testing these models on both CT images alone and fused PET and CT images offers a comprehensive evaluation, allowing us to demonstrate the advantages of our multi-modal approach over single-modality methods.

\subsection{Multi-modal Classification Model}
For the multi-modal classification model, we introduced two sophisticated model architectures: MedClip and BEiT (Bidirectional Encoder representation from Image Transformers). 
MedClip employs a dual encoder structure for visual and textual data, using cross-modal contrastive learning to align these representations in a unified feature space. This approach is particularly novel and crucial as it facilitates the seamless integration of medical images with clinical and genetic data, enhancing the model's ability to draw comprehensive insights from diverse data types \cite{wang2022medclip}. The dual encoder structure allows MedClip to effectively capture the complementary information from both image and text data, making it a powerful tool for multi-modal medical diagnostics.
BEiT, based on the Vision Transformer (ViT) architecture, leverages bidirectional context modeling and masked image modeling to develop robust representations from images. This model is pre-trained on medical images, which enhances its capability to detect and classify tumors accurately \cite{bao2021beit}. The use of bidirectional context modeling allows BEiT to consider the entire context of an image, providing a more thorough understanding of the visual information. Masked image modeling, on the other hand, trains the model to predict missing parts of the image, which improves its ability to generalize and perform well on unseen data. 
These models were tested with three different image input combinations: CT alone, CT and PET separately, and the fused CT and PET image, allowing us to evaluate the efficacy of our fusion method thoroughly. By doing so, we can determine the effectiveness of the proposed method and fusion model. The architecture of the best multi-modal model BEiT-based is shown in Figure \ref{beit}.

\section{Experimental Details}
\label{sec:typestyle}
\subsection{Datasets}
This paper utilized three distinct datasets to optimize the analysis of NSCLC. The primary dataset was the NSCLC Radiogenomic collection from the Cancer Imaging Archive \cite{bakr2018radiogenomic}. This comprehensive dataset includes 285,411 scan images from 303 studies of 211 NSCLC patients, encompassing both CT and PET lung images. It also provides extensive clinical data, covering variables such as age, smoking status, histology, treatment history, and cancer recurrence. Additionally, it includes RNA sequencing data from biopsied tumor tissues, linking genetic information with imaging.

Given the primary dataset's imbalance, predominantly featuring the Adenocarcinoma class, it was supplemented with the NSCLC Radiomics dataset \cite{aerts2015data}, which contains images of 422 NSCLC patients. Notably, both datasets were generated using the same scanner, ensuring consistency in data quality. The Radiomics dataset was exclusively used to augment the training data, while validation was performed solely on the primary dataset. The third dataset employed was a Large-Scale CT and PET dataset \cite{li2020large}, featuring de-noised PET scans. These scans were used to train a deep CNN auto-encoder with added noise to enhance the model's robustness. Post-training, this model was applied to our primary dataset for PET scan de-noising, thereby improving the quality of our input data for subsequent analysis.

\subsection{Implementation Detail and Result Analysis}
For the hardware requirement, computers equipped with NVIDIA Quadro RTX6000 were used to train the models, and the multi-core processors were used for efficient data processing and analysis. Powerful GPUs are necessary to speed up the computations significantly. Finally, given the large size of medical imaging datasets, storage solutions like cloud storage were used. In terms of software, 3D Slicer was used to view and pair the CT and PET DICOM scans and to register the image between the two scans. Moreover, several Python libraries were needed to aid in working with imaging scans, namely pydicom, pynrrd, skimage, lungmask \cite{Hofmanninger2020}, and SimpleITK, to name a few. Besides these, we used deep learning frameworks such as PyTorch, TensorFlow, and Keras to develop and train the models for image classification. Also, the library Optuna was used to facilitate the hyperparameters search \cite{akiba2019optuna}.

\subsection{Model Training}
For the deep learning models, the key hyperparameters included a learning rate of 0.001, a batch size of 96, a dropout of 0.5, and the Adam optimizer held constant across all models for fair comparison. For the XGBoost model for feature selection, the learning rate was set to 0.1, with a maximum depth of 5 for trees and 100 estimators. We employed a 5-fold cross-validation approach for testing and model evaluation. The baseline methods included a standard logistic regression model and support vector machine for tabular data comparison and several image classifiers such as 2D CNN, 3D CNN, VGG16, ResNet, Inception, and Xception. The optimal parameters for the tabular baseline models were achieved through GridSearch. The image baseline models were fed the CT images alone as well as the fused CT and PET, serving as a comparison point to assess the performance improvements offered by the more complex multi-modal models proposed. The assessment metrics focused on accuracy, precision, recall, and f1-score. 

\subsection{Evaluation Metrics}
For the classification task, we prioritized evaluation metrics crucial for medical diagnostics: accuracy, precision, recall, and F1-score. Accuracy measures the overall correctness of the model, while precision assesses the proportion of true positives among positive predictions. Recall, or sensitivity, quantifies the model's ability to identify true positives correctly. The F1-score, a balance of precision and recall, provides a holistic view of the model's performance. Additionally, we analyzed the confusion matrix to scrutinize false positives and negatives, which are particularly critical in medical applications. These metrics were integral not only for assessing model performance but also for comparing the effectiveness of different imaging inputs, such as standalone CT scans versus integrated fused scans.
\section{Results and Discussion}
\label{sec:majhead}
\subsection{Quantitative Analysis}
Our experimental analysis comprehensively evaluated various machine learning models, spanning tabular, image-based, and multi-modal approaches both separately and in a fused imaging modality. Key findings, detailed in Table 1, reveal distinct performance patterns across these models, evaluated on accuracy, precision, recall, and F1-score.

Baseline tabular models, specifically SVM and Logistic Regression, exhibited solid performance with an accuracy of 77.0\%. Image classifiers using solely CT scans showed varied effectiveness; 2D CNNs achieved a 70.0\% accuracy, whereas Inception models reached up to 79.0\%. Notably, the incorporation of fused CT and PET scans markedly boosted model performance. For instance, VGG16, using fused images, attained an 87.1\% accuracy and an 82.5\% F1-score, underlining the value of fused imaging.

The standout results were observed in multi-modal models. The MedClip model, integrating tabular, genetic, and fused imaging data, achieved an 84.9\% accuracy. The BEiT-based model, also utilizing a multi-modal approach, demonstrated a remarkable accuracy of 94.04\% with fused data. This represents a significant advancement over the performances achieved using separate CT and PET scans, underscoring the efficacy of our multi-modal, fused imaging methodology. These results not only highlight the potential of advanced AI models in medical diagnostics but also emphasize the transformative impact of integrating diverse data modalities.
\setlength\tabcolsep{2pt}
\begin{table}[!ht]
    \centering
    \begin{tabular}{lcccc}
        \toprule
         \thead{Model} &  \thead{Accuracy \\(\%)} &  \thead{Precision \\(\%)} &  \thead{Recall \\(\%)} &  \thead{F1-Score \\(\%)} \\
        \midrule
        \multicolumn{5}{c}{Tabular Models} \\
        \midrule
        SVM & \textbf{77.0 ± 1.7} & \text{76.5 ± 1.3} & \textbf{76.5 ± 1.8} & \textbf{76.0 ± 1.6} \\
        Logistic Reg. & \textbf{77.0 ± 1.6} & \textbf{79.5 ± 1.6} & \text{74.5 ± 1.9} & \text{75.0 ± 1.7} \\
        \midrule
        \multicolumn{5}{c}{Image Classifier with CT images} \\
        \midrule
        2D CNN & \text{70.0 ± 2.7} & \text{64.0 ± 1.8} & \text{64.5 ± 2.1} & \text{64.0 ± 2.2} \\
        3D CNN & \text{55.5 ± 2.8} & \text{44.0 ± 3.1} & \text{48.5 ± 2.8} & \text{40.0 ± 2.9} \\
        VGG16 & \text{75.0 ± 0.7} & \text{70.0 ± 0.9} & \text{71.5 ± 1.1} & \text{70.5 ± 0.5} \\
        ResNet & \text{74.0 ± 1.9} & \text{74.1 ± 2.6} & \textbf{73.7 ± 1.7} & \textbf{71.6 ± 1.8}\\
        Inception & \textbf{79.0 ± 1.3} & \textbf{76.0 ± 1.0} & \text{68.0 ± 0.8} & \text{70.0 ± 0.9} \\
        Xception & \text{78.0 ± 1.2} & \text{74.0 ± 0.8} & \text{70.0 ± 0.6} & \text{71.0 ± 0.7} \\
        \midrule
        \multicolumn{5}{c}{Image Classifier with Fused CT + PET images} \\
        \midrule
        2D CNN &  \text{75.0 ± 2.1} & \text{73.0 ± 2.6} &  \text{65.0 ± 2.2} & \text{66.5 ± 2.4} \\
        3D CNN & \text{60.0 ± 2.6} & \text{79.0 ± 2.3} & \text{55.0 ± 2.4} & \text{45.0 ± 2.5} \\
        VGG16 &\textbf{87.1 ± 0.8} & \text{83.0 ± 2.1} & \textbf{82.5 ± 1.9} & \textbf{82.5 ± 2.0} \\
        ResNet &  \text{82.0 ± 1.5} & \textbf{86.8 ± 2.2} & \text{81.9 ± 1.5} &  \text{81.3 ± 1.3}\\
        Inception & \text{81.8 ± 0.9} & \text{86.5 ± 1.2} & \text{82.0 ± 1.1} & \text{81.0 ± 0.8} \\
        Xception & \text{81.8 ± 1.1} & \text{86.5 ± 1.0} & \text{82.0 ± 0.9} & \text{81.0 ± 1.1} \\
        \midrule
        \multicolumn{5}{c}{Multi-modal MedClip Model (CT+PET+EHR+Genetic)} \\
        \midrule
         CT only &  \text{78.2 ± 1.9} & \text{79.0 ± 1.8} & \text{77.5 ± 2.3} & \text{78.0 ± 2.1} \\
        Separate CT/PET & \text{76.0 ± 1.7} & \text{76.0 ± 2.1} & \text{75.5 ± 1.3} & \text{75.5 ± 1.5} \\
        Fused CT/PET &  \textbf{84.9 ± 0.9} & \textbf{89.5 ± 1.1} & \textbf{83.0 ± 1.3} & \textbf{83.5 ± 1.5} \\
        \midrule
        \multicolumn{5}{c}{Multi-modal BEiT Model (CT+PET+EHR+Genetic)} \\
        \midrule
        CT only & \text{85.6 ± 1.7} & \text{85.5 ± 2.1} & \text{85.5 ± 2.3} & \text{85.5 ± 2.6} \\
        Separate CT/PET & \text{76.0 ± 1.4} & \text{76.0 ± 2.5}  & \text{75.5 ± 1.7} &  \text{75.5 ± 1.9}\\
        Fused CT/PET  & \textbf{94.0 ± 0.7} & \textbf{95.0 ± 1.2} & \textbf{94.0 ± 1.0} & \textbf{94.0 ± 1.1} \\
        \bottomrule
    \end{tabular}
    \caption{Comparison of all models evaluated with varied input of CT alone versus fused CT/PET.}
    \label{tab:my_label}
\end{table}

\begin{figure}[!htp]
    \centering
\includegraphics[width=0.45\textwidth]{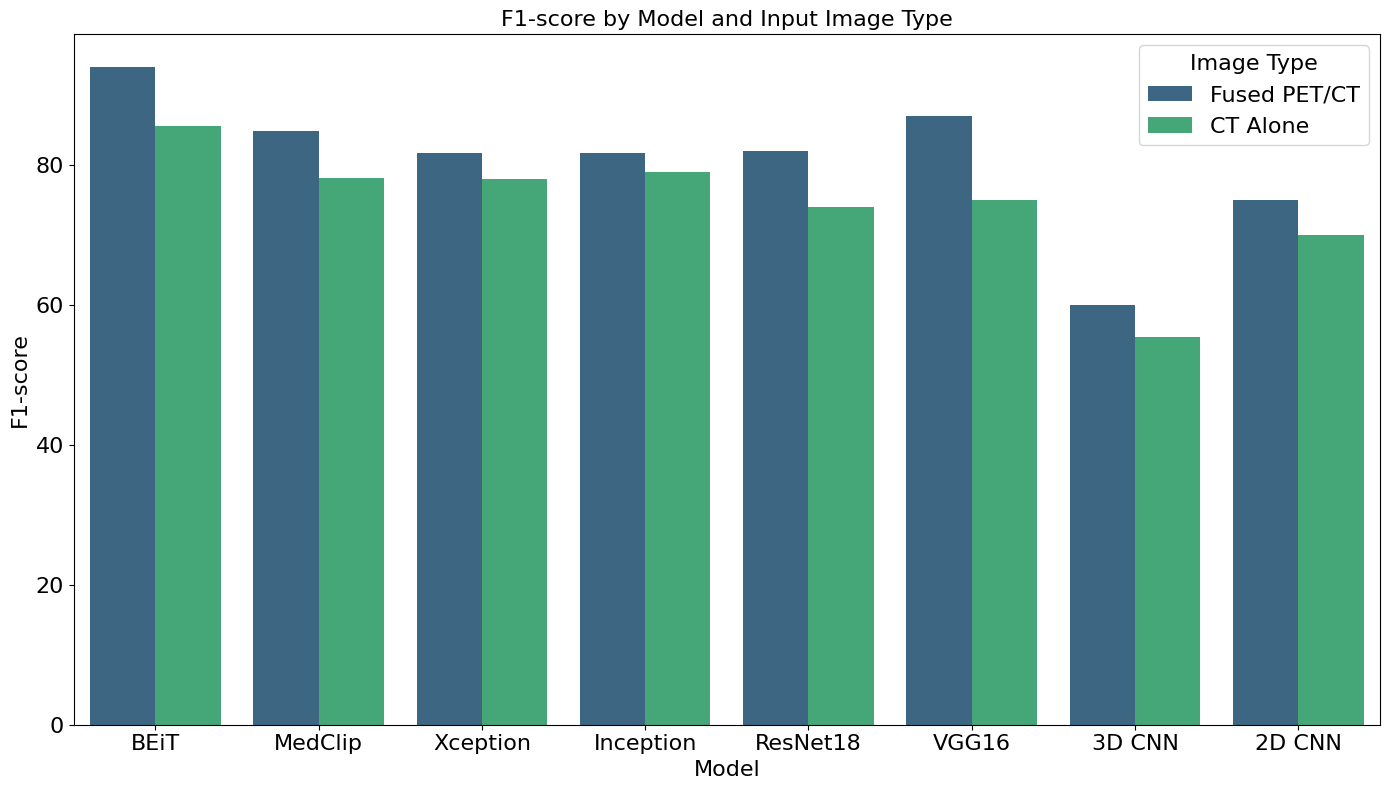}
    \caption{F1-score comparison of different models using fused PET/CT images versus CT images alone.}
    \label{fig:f1_models}
\end{figure}

\subsection{Qualitative Analysis}
Throughout all the assessments, the implementation of fused imaging data notably improved model performance metrics across the board, underscoring the value of integrating multiple imaging modalities for enhanced diagnostic inference, as illustrated in Figure \ref{fig:f1_models}. The deep learning models, particularly the one based on the BEiT architecture, effectively leveraged the rich, multi-modal data, translating into superior quantitative outcomes. It is also important to note that the multi-modal model surpasses the imaging and tabular model in performance, justifying the need to integrate data from different modalities to enhance the diagnosis accuracy. Moreover, even when trying the same multi-modal model but processing the CT and PET scans separately through feature extraction and then combining the results always performed worse than using the fused image generated by the VGG19 fusion model, as it can be concluded that the fusion process is necessary and aids in the improved performance of the overall model. Regarding the pre-processing of the imaging scans, the deep CNN auto-encoder helped a lot in de-noising the PET scan, and the results were very noticeable. Similarly, for the CT scans, the pre-processing was applied to filter only the slices containing lung pixels, and masking and improving the contrast helped improve the results.
\subsection{Discussion}
The quantitative leap in performance metrics with fused imaging data suggests that models benefit from the complementary information available in different imaging modalities. It can be inferred that the spatial resolution of CT images combined with the metabolic information from PET scans provides a more holistic view of the pathology, which is effectively exploited by the more complex architectures like VGG16, Inception, and the BEiT models. This illustrates that although the BEit model uses fewer parameters than MedClip and VGG16, it had a better performance. 

The SVM and Logistic Regression models, while less effective than image classifiers, offer competitive performance on tabular data, indicating that traditional machine learning models remain valuable for structured data analysis. The lower performance of 3D CNNs across all metrics suggests that for the dataset at hand, the additional complexity introduced by 3D convolution may not capture the essential features as effectively as other architectures, possibly due to overfitting or the need for more training data.

The superior results of the BEiT-based model highlight the potential of transformer architectures in handling complex, multi-modal datasets. Transformers’ ability to model long-range dependencies and integrate disparate data sources is particularly beneficial for medical imaging tasks, where the context and subtlety of features are crucial for accurate diagnosis.

\section{Limitations}
\label{s:limitations}
The primary limitation of our multi-modal approach is the constrained dataset size, as comprehensive data across all modalities (clinical, imaging, and genomic) was available for only a limited subset of patients. This reduction in dataset size potentially affects the generalizability and robustness of our models and necessitates the use of supplemental data from a secondary source to address class imbalances. Additionally, the complexity of integrating diverse data sources, while beneficial for performance, leads to the 'black box' issue, diminishing the interpretability of the model's decision-making process. This aspect is particularly critical in clinical settings where transparency is essential for clinician trust and model applicability. Future efforts might focus on enhancing model transparency through explainable AI techniques.
\section{Conclusion}
\label{s:conclusion}
In conclusion, this paper made significant strides in NSCLC diagnostics by integrating fused CT and PET scans with genomics and clinical data. Our key contributions include the development of a novel image fusion technique that combines anatomical and metabolic information, the innovative application of advanced models like MedClip and BEiT for multi-modal data analysis, and the implementation of sophisticated pre-processing and denoising techniques for CT and PET scans. These contributions led to a substantial improvement in NSCLC classification, achieving a remarkable accuracy of 94.04\%. This success demonstrates the transformative potential of integrating diverse data sources and leveraging state-of-the-art transformer-based architectures in medical diagnostics. The high accuracy achieved underscores the critical importance of fused imaging and comprehensive data integration in providing a more detailed and accurate lung cancer analysis. These findings mark a significant leap forward in the field, setting a new standard for future research to develop even more sophisticated models and richer datasets, ultimately leading to more precise and reliable diagnostic tools in oncology.

\textbf{\\Future Work.} Future work will focus on expanding data sources by incorporating additional modalities such as proteomics and metabolomics, conducting longitudinal studies to track tumor progression, and validating model performance through real-world clinical trials. Enhancing model interpretability for clinical use, integrating our approach into existing clinical workflows, and optimizing for scalability and computational efficiency are also key areas for further research. These efforts aim to build on our findings to create more robust, accurate, and clinically applicable diagnostic tools. The ultimate goal is to improve patient outcomes in NSCLC and other cancers, a significant and impactful direction for future research.

\bibliographystyle{IEEEbib}
\bibliography{strings,refs}

\begin{thebibliography}{10}

\bibitem{WHO}
WHO,
\newblock ``World health organization: Lung cancer fact sheet,'' https://www.who.int/news-room/fact-sheets/detail/lung-cancer.

\bibitem{dodia2022recent}
Shubham Dodia, B~Annappa, and Padukudru~A Mahesh,
\newblock ``Recent advancements in deep learning based lung cancer detection: A systematic review,''
\newblock {\em Engineering Applications of Artificial Intelligence}, vol. 116, pp. 105490, 2022.

\bibitem{naseer2023lung}
Iftikhar Naseer, Tehreem Masood, Sheeraz Akram, Arfan Jaffar, Muhammad Rashid, and Muhammad~Amjad Iqbal,
\newblock ``Lung cancer detection using modified alexnet architecture and support vector machine.,''
\newblock {\em Computers, Materials \& Continua}, vol. 74, no. 1, 2023.

\bibitem{chaunzwa2021deep}
Tafadzwa~L Chaunzwa, Ahmed Hosny, Yiwen Xu, Andrea Shafer, Nancy Diao, Michael Lanuti, David~C Christiani, Raymond~H Mak, and Hugo~JWL Aerts,
\newblock ``Deep learning classification of lung cancer histology using ct images,''
\newblock {\em Scientific reports}, vol. 11, no. 1, pp. 1--12, 2021.

\bibitem{amini2021multi}
Mehdi Amini, Mostafa Nazari, Isaac Shiri, Ghasem Hajianfar, Mohammad~Reza Deevband, Hamid Abdollahi, Hossein Arabi, Arman Rahmim, and Habib Zaidi,
\newblock ``Multi-level multi-modality (pet and ct) fusion radiomics: prognostic modeling for non-small cell lung carcinoma,''
\newblock {\em Physics in Medicine \& Biology}, vol. 66, no. 20, pp. 205017, 2021.

\bibitem{alduais2023non}
Yaser Alduais, Haijun Zhang, Fan Fan, Jing Chen, and Baoan Chen,
\newblock ``Non-small cell lung cancer (nsclc): A review of risk factors, diagnosis, and treatment,''
\newblock {\em Medicine}, vol. 102, no. 8, pp. e32899, 2023.

\bibitem{feng2022deep}
Jianxin Feng, Jun Jiang, et~al.,
\newblock ``Deep learning-based chest ct image features in diagnosis of lung cancer,''
\newblock {\em Computational and Mathematical Methods in Medicine}, vol. 2022, 2022.

\bibitem{khoirunnisa2023implementation}
Azka Khoirunnisa, Didit Adytia, et~al.,
\newblock ``Implementation of crnn method for lung cancer detection based on microarray data,''
\newblock {\em JOIV: International Journal on Informatics Visualization}, vol. 7, no. 2, pp. 600--605, 2023.

\bibitem{zhang2015relationship}
Jianjun Zhang, Kathryn~A Gold, Heather~Y Lin, Stephen~G Swisher, Yan Xing, J~Jack Lee, Edward~S Kim, and William~N William~Jr,
\newblock ``Relationship between tumor size and survival in non--small-cell lung cancer (nsclc): an analysis of the surveillance, epidemiology, and end results (seer) registry,''
\newblock {\em Journal of Thoracic Oncology}, vol. 10, no. 4, pp. 682--690, 2015.

\bibitem{han2021histologic}
Yong Han, Yuan Ma, Zhiyuan Wu, Feng Zhang, Deqiang Zheng, Xiangtong Liu, Lixin Tao, Zhigang Liang, Zhi Yang, Xia Li, et~al.,
\newblock ``Histologic subtype classification of non-small cell lung cancer using pet/ct images,''
\newblock {\em European journal of nuclear medicine and molecular imaging}, vol. 48, pp. 350--360, 2021.

\bibitem{paudyal2023artificial}
Ramesh Paudyal, Akash~D Shah, Oguz Akin, Richard~KG Do, Amaresha~Shridhar Konar, Vaios Hatzoglou, Usman Mahmood, Nancy Lee, Richard~J Wong, Suchandrima Banerjee, et~al.,
\newblock ``Artificial intelligence in ct and mr imaging for oncological applications,''
\newblock {\em Cancers}, vol. 15, no. 9, pp. 2573, 2023.

\bibitem{wang2022medclip}
Zifeng Wang, Zhenbang Wu, Dinesh Agarwal, and Jimeng Sun,
\newblock ``Medclip: Contrastive learning from unpaired medical images and text,''
\newblock {\em arXiv preprint arXiv:2210.10163}, 2022.

\bibitem{bao2021beit}
Hangbo Bao, Li~Dong, Songhao Piao, and Furu Wei,
\newblock ``Beit: Bert pre-training of image transformers,''
\newblock {\em arXiv preprint arXiv:2106.08254}, 2021.

\bibitem{bakr2018radiogenomic}
Shaimaa Bakr, Olivier Gevaert, Sebastian Echegaray, Kelsey Ayers, Mu~Zhou, Majid Shafiq, Hong Zheng, Jalen~Anthony Benson, Weiruo Zhang, Ann~NC Leung, et~al.,
\newblock ``A radiogenomic dataset of non-small cell lung cancer,''
\newblock {\em Scientific data}, vol. 5, no. 1, pp. 1--9, 2018.

\bibitem{aerts2015data}
HJWL Aerts, E~Rios Velazquez, RT~Leijenaar, Chintan Parmar, Patrick Grossmann, S~Cavalho, Johan Bussink, Ren{\'e} Monshouwer, Benjamin Haibe-Kains, Derek Rietveld, et~al.,
\newblock ``Data from nsclc-radiomics,''
\newblock {\em The cancer imaging archive}, 2015.

\bibitem{li2020large}
Ping Li, S~Wang, T~Li, J~Lu, Y~HuangFu, and D~Wang,
\newblock ``A large-scale ct and pet/ct dataset for lung cancer diagnosis [dataset],''
\newblock {\em The cancer imaging archive}, 2020.

\bibitem{Hofmanninger2020}
Johannes Hofmanninger, Forian Prayer, Jeanny Pan, Sebastian R\"{o}hrich, Helmut Prosch, and Georg Langs,
\newblock ``Automatic lung segmentation in routine imaging is primarily a data diversity problem, not a methodology problem,''
\newblock {\em European Radiology Experimental}, vol. 4, no. 1, Aug. 2020.

\bibitem{akiba2019optuna}
Takuya Akiba, Shotaro Sano, Toshihiko Yanase, Takeru Ohta, and Masanori Koyama,
\newblock ``Optuna: A next-generation hyperparameter optimization framework,''
\newblock in {\em Proceedings of the 25th ACM SIGKDD international conference on knowledge discovery \& data mining}, 2019, pp. 2623--2631.

\end{thebibliography}

\end{document}